\documentclass[12pt,a4paper]{article}

\usepackage{setspace}
\usepackage{a4wide}
\usepackage{amssymb}
\usepackage{amsmath}
\usepackage{graphicx}
\usepackage{cancel}
\usepackage{array}
\usepackage{slashed}
\usepackage{epsfig} 
\usepackage{hyperref}
\newcolumntype{C}{>{\centering\arraybackslash}m{4cm}}
\newcommand{\beq}{\begin{equation}}
\newcommand{\eeq}{\end{equation}}
\newcommand{\beqa}{\begin{eqnarray}}
\newcommand{\eeqa}{\end{eqnarray}}

\newcommand{\ft}[2]{{\textstyle\frac{#1}{#2}}}

\begin{document}
\onehalfspace

%%%%%%%%%%%%%%%%%%%%%%%%%%%%%%%%%%%%%%%%%%%%%%%%%%%%%%%%%%%%%%%%%%%%%%

\thispagestyle{empty}

\begin{titlepage}
\begin{flushright}\footnotesize
\vspace{0.5cm}
\end{flushright}
\setcounter{footnote}{0}

\begin{center}
{\Large{\bf Scaling dimensions at small spin in $\mathcal{N}=4$ SYM theory}}
\vspace{15mm}

\centerline{\sc B.~Basso\footnote{Email: bbasso@princeton.edu}}

\vspace{10mm}

\centerline{\it Princeton Center for Theoretical Science,}
 \centerline{\it Jadwin Hall, Princeton University,}
\centerline{\it NJ 08544, USA}

\vspace{1cm}

\textbf{Abstract}

\vspace{5mm}

\end{center}

\noindent{In this paper, we examine scaling dimensions at small spin in the so-called $\mathfrak{sl}(2)$ sector of the planar maximally supersymmetric Yang-Mills ($\mathcal{N}=4$ SYM) theory. We find that the Bethe ansatz equations, which control the spectrum of scaling dimensions, greatly simplify in this limit and can be solved exactly.}

\end{titlepage}

\newpage

\setcounter{footnote} 0

{\small \tableofcontents}

\newpage

%%%%%%%%%%%%%%%%%%%%%%%%%%%%%%%%%%%%%%%%%%%%%%%%%%%%%%%%%%%%%%%%%%%%%%

\section{Introduction}

In this paper, we consider the spectrum of scaling dimensions in the so-called $\mathfrak{sl}(2)$ sector of planar $\mathcal{N}=4$ SYM theory. More precisely, we shall consider local single-trace operators,
\beq
\mathcal{O} = \textrm{tr}\, D^{k_{1}}Z\ldots D^{k_{J}}Z\, ,
\eeq
that are made out of $J$ complex scalar fields $Z$ and $k_{1} + \ldots + k_{J} = S$ light-cone covariant derivatives $D$. Scaling dimensions $\Delta \equiv \Delta_{J, S}(\lambda)$ in this sector are in general complicated functions of the spin $S$, twist $J$, and 't Hooft coupling constant $\lambda \equiv g^2_{YM}N_{c}$. Not to mention their dependence on the `hidden' quantum numbers -- mode numbers -- that control the fine structure of the spectrum.

A powerful tool for computing scaling dimensions in planar $\mathcal{N}=4$ SYM theory is integrability (see~\cite{Review} for a recent review). In this framework, scaling dimensions are determined by a set of non-linear integral equations, known as TBA/Y-system equations~\cite{GKV, AF}. The seed for solving these equations comes from the Asymptotic Bethe Ansatz (ABA) equations~\cite{BS05, BES}. In some cases, the prediction from the ABA equations is already complete and receives no extra corrections, which are otherwise known as wrapping corrections~\cite{AJK,KLRSV}. This is well-illustrated by the large spin limit of the scaling dimensions, which has remarkable properties~\cite{BGK} and for which an exact integral equation was derived directly from ABA data~\cite{BES}. A second example is the small spin limit, where wrapping corrections appear miraculously suppressed.

The small spin limit of scaling dimensions was first considered in~\cite{KLOV}, at twist two, and later in~\cite{Slope}, at higher twist. They were found to admit the small spin expansion
\beq\label{DeltaIntro}
\Delta = J + \alpha_{J}(\lambda) S + O(S^2)\, ,
\eeq
which implicitely requires an extrapolation from integer to continuous spin $S$. Under this assumption, it was observed~\cite{Slope}, from explicitely known weak coupling expressions~\cite{KLOV, KLRSV}, that the first Taylor coefficient $\alpha_{J}(\lambda)$ in~(\ref{DeltaIntro}) can be found exactly for the minimal scaling dimensions (i.e., these with lowest mode numbers). This quantity does not apparently receive wrapping corrections~\cite{Slope, GV} and can be written in closed form as~\cite{Slope}
\beq\label{slope}
\alpha_{J}(\lambda) = 1+{\sqrt{\lambda} \over J}{I_{J+1}(\sqrt{\lambda}) \over I_{J}(\sqrt{\lambda})}\, ,
\eeq
where $I_{k}(x)$ is the $k$-th Bessel's function. This expression passed furthermore non-trivial tests at strong coupling~\cite{GV}. It was also used as an input for a computation of the Konishi anomalous dimension at two-loop at strong coupling~\cite{GV} -- leading to a proposal which appears consistent with both numerics from TBA equations~\cite{GV, Frolov} and (partial) semiclassical string calculation~\cite{BGMRT}.

In this paper, we shall derive the formula~(\ref{slope}) directly from the ABA equations for planar $\mathcal{N}=4$ SYM theory~\cite{BS05, BES}. More precisely, we shall solve the long-range Baxter equation~\cite{Bel06, Bel09}, whose dynamical content is equivalent to the one of the ABA equations.

We shall see moreover that the formula~(\ref{slope}) admits a generalization to some non-minimal scaling dimensions (i.e., those with higher mode numbers).

The absence of wrapping corrections at small spin remains surprising and will not be elucidated here. This feature seems moreover to be specific to planar $\mathcal{N}=4$ SYM theory. For comparison, in the related context of the planar ABJM theory, the slope of the minimal scaling dimensions is consistent with
\beq\label{ABJM}
\alpha_{J}^{ABJM}(\lambda) = 1 + {2\pi h(\lambda) \over J}{I_{J+1}(2\pi h(\lambda)) \over I_{J}(2\pi h(\lambda))}\, ,
\eeq
with $h(\lambda) = h(N_{c}/k)$, the interpolating function of the ABA equations for ABJM theory~\cite{GVaba} (see also references therein).%
\footnote{No derivation of the formula~(\ref{ABJM}) will be proposed here. It is just observed to be consistent with explicit ABA results~\cite{Z,BM} at weak coupling for twist $J=1$ and (minimal) $J=2$ -- assuming the choice of the even-spin trajectory, as otherwise the scaling dimension will not even vanish at small spin after continuation.}
But, as opposed to what happens in planar $\mathcal{N}=4$ SYM theory and even if~(\ref{ABJM}) were confirmed as a prediction of the ABA equations, it could not stand as an exact result for planar ABJM theory. Indeed, wrapping corrections do contribute at small spin in the ABJM theory and should be added to~(\ref{ABJM}). This is visible at both weak and strong coupling~\cite{Beccaria}.

The plan of the paper is the following. In Section~2, we start with some preliminary comments and present our general expression for scaling dimensions, and higher conserved charges, at small spin. In Section~3, we present our strategy for solving the long-range Baxter equation at small spin, and, in Section~4, we provide its general solution. The technical details of the analysis are given in the Appendix.

\section{Preliminary comments}

A solution to the ABA equations is given once a set of $S$ roots $u_{k}$ -- the Bethe roots -- is known.  One then determines the so-called higher conserved charges $Q_{n+1}$ by means of
\beq\label{QMR}
Q_{n+1} = {1\over n} \sum_{k=1}^{S}\left({i \over x^{+n}_k} - {i \over x^{-n}_k}\right)\, ,
\eeq
where $x^{\pm n}_k \equiv (x^{\pm}_k)^n$, $x^{\pm}_k \equiv x(u_k\pm i/2)$, $x(u) \equiv (u+\sqrt{u^2-(2g)^2})/2$ and $g \equiv \sqrt{\lambda}/4\pi$.%
\footnote{In the following, it will be useful to consider $n\in \mathbb{Z}$, thus enlarging the standard definition of the charges to negative integers as well. The case $n=0$ is obtained as a limit, and the associated charge, known as the total momentum, reads
\beq
Q_{1} \equiv -i\sum_{k=1}^{S}\log{\left({x^{+}_{k} \over x^{-}_{k}}\right)}\, .
\eeq
}
From them, the scaling dimension follows,%
\footnote{Up to wrapping corrections.}
\beq\label{DeltaQ2}
\Delta = J + S + 2g^2Q_{2}\, .
\eeq
In this paper, we consider the possibility to extrapolate at small spin the computation of scaling dimensions. This calls for a generalization of the ABA equations to deal with non-integer spin $S$, i.e., with solutions which are not characterized by a set of Bethe roots. Our main hypothesis is that this extension, which is not free of ambiguity, is not void of any sense.

Technically, we assume that the charges have a smooth limit at small spin. Namely, for our analysis to apply, they should admit the expansion
\beq\label{Smooth}
Q_{n+1} = q_{n+1} S + O(S^2)\, ,
\eeq
for $S \sim 0$. This relation puts as the restriction that $Q_{n+1} (S=0)  = 0$, which characterizes the (BPS) vacuum. In other words, trajectories that do not satisfy the latter relation, i.e., for which $Q_{n+1} (S=0) \neq 0$, are not covered by our analysis. In the other cases, the rate at which the charges vanish is controlled by the reduced charges $q_{n+1}$, which contain both the coupling constant dependence and the information on the state under study. 

Implementing the condition~(\ref{Smooth}), and the small spin expansion in general, can be performed in the framework of the Baxter equation (see next Section). It is indeed easier to eliminate explicit reference to the Bethe roots (and then to their inherently discrete nature) at the level of the latter equation. In other words, the constraint of integer spin is more easily relaxed. One can then solve directly for the charges $Q_{n+1}$ and find solutions satisfying~(\ref{Smooth}).

Moreover, the Baxter equation simplifies drastically in the small spin limit and solutions of the type~(\ref{Smooth}) can be constructed exactly. One finds an (elementary) solution with charges $q_{n+1}^{(m)}$ for any mode number $m \in \mathbb{Z}^*$.%
\footnote{A solution at mode zero $m=0$ can be found but is somewhat ambiguous.} The latter parameter can be read directly from the first charge
\beq
q^{(m)}_{1} = {2\pi m\over J}\, .
\eeq
A general solution is then obtained by summing over such elementary solutions,
\beq\label{SMN}
q_{n+1} = \sum_{m} \kappa_{m} q_{n+1}^{(m)}\, ,
\eeq
where the arbitrary cofficients $\kappa_{m}$ are the filling fractions satisfying $\sum_{m} \kappa_{m}=1$.%
\footnote{Our analysis requires furthermore that the total charge $Q_{1} = Sq_{1} = S\sum_{m}\kappa_{m} q_{1}^{(m)}$ vanishes.}

Our main result is the expression for the elementary charges $q^{(m)}_{n+1}$ at mode number $m$. The are found to admit the representation
\beq\label{ResInt}
2g^{n+1} q^{(m)}_{n+1} = {m\sqrt{\lambda}\over J} {I_{J+n}(m\sqrt{\lambda}) \over I_{J}(m\sqrt{\lambda})} \, ,
\eeq
where $\lambda \equiv 4\pi g$ is the 't Hooft coupling and $I_{k}(x)$ is the $k$-th Bessel's function, with the small $x$ behavior $I_{k}(x) \sim x^k/2^k\Gamma(k+1)$.

In the particular case of the twist $J$ minimal scaling dimension only the two mode numbers $m=\pm 1$ are filled, with equal filling fractions $\kappa_{1} = \kappa_{-1} = 1/2$. As an immediate consequence of~(\ref{Smooth}, \ref{SMN}, \ref{ResInt}), the scaling dimension~(\ref{DeltaQ2}) is found to be of the form~(\ref{DeltaIntro}), with the slope $\alpha_{J}(\lambda) = 1+g^2(q^{(1)}_{2} + q^{(-1)}_{2})$ reproducing exactly the expression~(\ref{slope}) given in the introduction.

Our proposal~(\ref{ABJM}) for the ABJM theory relies on the expectation that the formula~(\ref{ResInt}) applies to this case as well, with, for the minimal scaling dimensions, the mode numbers $m=\pm 1/2$ equally filled this time. Together with the replacement rule $\sqrt{\lambda} \rightarrow 4\pi h(\lambda)$, this leads directly to~(\ref{ABJM}).  It would be interesting to test this expression using the long-range Baxter equation for the (pseudo-)$\mathfrak{sl}(2)$ sector of the ABJM theory. 

The rest of the paper provides the proof of~(\ref{ResInt}), using the long-range Baxter equation.

\section{Baxter equation}

The long-range Baxter equation can be found as%
\footnote{We mostly adapt from~\cite{BaBel}.}
\beq\label{Beq}
\Delta_{+}(u+\ft{i}{2})R(u+\ft{i}{2}) + \Delta_{-}(u-\ft{i}{2})R^{-1}(u-\ft{i}{2}) = t(u)\, ,
\eeq
if written for the ratio
\beq\label{Ru}
R(u) \equiv {Q(u+\ft{i}{2}) \over Q(u-\ft{i}{2})}\, ,
\eeq
with $Q(u)$ the Baxter $Q$-function. At integer spin $S$, the latter is given by a polynomial in the rapidity $u$ of degree $S$,
\beq
Q(u) = \prod_{j=1}^{S}(u-u_{j})\, ,
\eeq
whose roots $u_{j}$ are solutions to the ABA equations,
\beq
1+ {\Delta_{-}(u-i/2) Q(u-i) \over \Delta_{+}(u+i/2) Q(u+i)}= 0\, .
\eeq
In the $\mathfrak{sl}(2)$ sector, the Bethe roots are always real and the function $R(u)$ has poles and zeros outside of the real $u$-axis.

The other quantities involved in the Baxter equation~(\ref{Beq}) are the dressing factors $\Delta_{\pm}(u)$ and the $t$-function $t(u)$. Both are `dynamical' quantities, in the sense that they depend on the state under study. Their precise forms will not play an essential role in the following analysis. Technical details are given in Appendix~\ref{App}. Here we simply note their expressions in the limit where all charges vanish. They can be thought of as the values in the vacuum and read explicitely
\beq
\Delta^{\textrm{vac}}_{+}(u) = \Delta^{\textrm{vac}}_{-}(u) = x^J\, , \qquad t_{\textrm{vac}}(u) = x^{+J} + x^{-J}\, ,
\eeq
where $J$ is the twist. They are obtained by plugging the vacuum solution $Q(u)=1$ everywhere it appears in the Baxter equation~(\ref{Beq}).

\subsection{Two convenient parameterizations}

Prior to analyze the small spin limit of the Baxter equation~(\ref{Beq}) and its solutions, we first introduce two convenient parameterizations for the function $R(u)$. They can be seen as small and large $u$ parameterization, respectively.
 
For a polynomial solution, with $S$ Bethe roots $u_{j}$, we can write the function $R(u)$ in the form
\beq
R(u) =  \prod_{j=1}^{S}{u-u_j+\ft{i}{2} \over u-u_j-\ft{i}{2}} =  \prod_{j=1}^{S}{x-x^-_j \over x-x^+_j}{1-g^2/xx^-_j \over 1-g^2/xx^+_j}\, ,
\eeq
where in the last equality we applied the identity $u\pm i/2 = x^{\pm} + g^2/x^{\pm}$. We can now think of this ratio in two different ways, depending on whether we are interested in the large or small rapidity behavior. 

In the first case, we note that
\beq\label{RSmall}
R(u) = \exp{\left[-i\sum_{n\geq 1}Q_{n+1}x^n -i\sum_{n\geq 1}Q_{n+1}\left({g^2 \over x}\right)^n\right]}\, ,
\eeq
with the charges $Q_{n+1}$ as defined in~(\ref{QMR}) and where we used that $\prod_{j=1}^{S}x^{+}_j/x^{-}_j \equiv \exp{(iQ_{1})} =1$ for physical states. From the relation~(\ref{RSmall}), we learn that the charges $Q_{n+1}$ admit the two equivalent contour-integral representations
\beq\label{QRu}
Q_{n+1} ={1\over 2\pi}\oint {dx \over x^{n+1}}\log{R(u)} = {1\over 2\pi}\oint {dx\over x}\left({x \over g^2}\right)^{n}\log{R(u)}\, ,
\eeq
where, in both cases, the contour of integration encircles the interval $u^2 \leqslant (2g)^2$ counterclockwise. The equivalence of these two integral representations expresses the regularity of the function $\log{R(u)}$ around $u=0$, which reads
\beq\label{RC}
\oint {dx \over x}\left[x^n-\left({g^2 \over x}\right)^n\right]\log{R(u)} = 0\, ,
\eeq
for $n=1, 2, \ldots\, $.%
\footnote{The form~(\ref{RC}) of the regularity condition is convenient for our purposes, but equivalent to the more transparent one:
\beq
\oint {du \over u} u^{n}\log{R(u)} = 0\, ,
\eeq
for $n=1,2,  \ldots \, ,$ and with the integration performed around $u=0$.}

A second parameterization, which is more convenient at large rapidity, is found as
\beq\label{RLarge}
R(u) = \exp{\left[i\sum_{n\geq 1}Q_{-n+1}\left({1\over x}\right)^n -i\sum_{n\geq 1}Q_{n+1}\left({g^2 \over x}\right)^n\right]}\, ,
\eeq
where $Q_{-n+1}$ is nothing else than the continuation of $Q_{n+1}$ for negative values of $n$,
\beq\label{Qmin}
Q_{-n+1} = -{i\over n}\sum_{j=1}^{S}\left(x^{+n}_{j}-x^{-n}_{j}\right)\, .
\eeq
We note in particular that
\beq\label{Q0Q2}
Q_{0} = S + g^2Q_{2}\, ,
\eeq
which follows immediately from $u^{\pm} = x^{\pm}+g^2/x^{\pm}$.%
\footnote{The charge $Q_{0}$ is related to the renormalized $\mathfrak{sl}(2)$ conformal spin $j$, which reads $j = S + J/2+\gamma/2$, with $\gamma = 2g^2Q_{2}$ the anomalous dimension.} The identity~(\ref{Q0Q2}) should hold for arbitrary (positive) integer spin $S$. In the following, we shall assume that it holds true after continuation to any spin, and notably at small spin. It will allow us to apply the relation~(\ref{Q0Q2}) for normalizing our small spin solution.

As a side remark, we notice that the identity~(\ref{Q0Q2}) implies that the large $u$ asymptotics of $R(u)$, see~(\ref{RLarge}), satisfies
\beq\label{LUR}
\log{R(u)} = i(Q_{0}-g^2Q_{2})/u + O(1/u^2) = iS/u + O(1/u^2)\, .
\eeq
This is certainly what is expected when $Q(u) \sim u^S$.

\subsection{Linearized Baxter equation}

As alluded to before, our main assumption for constructing small spin solutions is that the charges $Q_{n+1}$ admit a regular expansion around $S=0$, starting as
\beq\label{WA}
Q_{n+1} = q_{n+1}S + O(S^2)\, .
\eeq
The coefficients $q_{n+1}$, with $n\in \mathbb{Z}$, are functions of the coupling $g$ and quantum numbers (e.g., twist $J$, mode numbers, ...) that specify the trajectory of scaling dimensions which we extrapolate down to small spin. The ansatz~(\ref{WA}) is consistent with the explicit form of the charges at arbitrary integer spin for twist two and (minimal) twist three, at weak coupling~\cite{BBKZ}. This observation motivates its consideration in the most general condition of our analysis.

Plugging the ansatz~(\ref{WA}) into the expression~(\ref{RSmall}), for the `generating' function $R(u)$, leads us to consider a solution to the Baxter equation~(\ref{Beq}) of the form
\beq\label{Rru}
R(u) = 1+r(u) S + O(S^2)\, ,
\eeq
where the function $r(u)$ is found as
\beq\label{SRSU}
r(u) =  -i\sum_{n\geq 1}q_{n+1}x^n - i\sum_{n\geq 1}q_{n+1}\left({g^2 \over x}\right)^n\, .
\eeq
As before, this representation tells us that the charges $q_{n+1}$ can be obtained from the contour integrals
\beq\label{Qru}
q_{n+1} ={1\over 2\pi}\oint {dx \over x^{n+1}} r(u) = {1\over 2\pi}\oint {dx\over x}\left({x \over g^2}\right)^{n} r(u)\, ,
\eeq
which both arrive from~(\ref{QRu}) after using~(\ref{Rru}). They enforce the regularity of $r(u)$ at $u=0$, 
\beq\label{regC}
{1\over 2i\pi}\oint {dx \over x}\bigg[x^{k}-\left({g^2 \over x}\right)^k\bigg] r(u)  = 0\, ,
\eeq
for $k=1, 2, \ldots\,$.

We now apply our small spin ansatz~(\ref{WA}, \ref{Rru}) to the Baxter equation~(\ref{Beq}). As shown in Appendix~\ref{App}, the Baxter equation~(\ref{Beq}) linearizes and can be written as
\beq\label{LBE}
x^{+J}r(u+\ft{i}{2}) - x^{-J}r(u-\ft{i}{2}) = t_{\textrm{eff}}(u)\, .
\eeq
Here $t_{\textrm{eff}}(u)$ is an effective $t$-function at small spin, which absorbs all source terms, coming both from the small spin expansion of the original $t$-function and from the dressing factors. It explicit form and relation to the original $t$-function are both given in Appendix~\ref{App}.

\subsection{Fixing the $t$-function}

An important remark concerning the effective $t$-function is that it contains a part which is polynomial in $u$ and which we denote as $\delta p(u)$. Namely,
\beq
t_{\textrm{eff}}(u) = \delta p(u) + \ldots\, ,
\eeq
where the dots stand for non-polynomial corrections (see Appendix~\ref{App}) and with $\delta p(u)$ of degree $J-1$ in $u$, at twist $J$. The latter polynomial has to be fixed appropriately if we want our problem to be related to the small spin limit of a polynomial solution. To do so, we linearize the large $u$ parameterization~(\ref{RLarge}), assuming the charges~(\ref{Qmin}) admit the expansion~(\ref{WA}) as well. It yields
\beq\label{rprim}
\hat{r}(u) = i\sum_{n\geq 1}q_{-n+1}\left({1\over x}\right)^n -i\sum_{n\geq 1}q_{n+1}\left({g^2 \over x}\right)^n\, ,
\eeq
as an ansatz. Notice that the function $\hat{r}(u)$ is not identical to the function $r(u)$ given in~(\ref{SRSU}), neither are these two functions asymptotic to one another at large $u$. We shall come back to this point later on.

We require now that $\hat{r}(u)$ solves the Baxter equation~(\ref{LBE}). As shown in Appendix~\ref{App},%
\footnote{For physical states only, i.e., such that $\exp{(iQ_{1})} =1$.} this can be done, if and only if
\beq\label{qqrel}
q_{-J-k+1} = g^{2k}q_{-J+k+1}\, ,
\eeq
for $k \neq J$ and $q_{-2J+1} = 0$. The polynomial part of the effective $t$-function is fixed along the way, as expected. For illustration, it reads
\beq
\delta p(u) = i\sum_{n=1}^{J} q_{-n+1} ((u+\ft{i}{2})^{J-n} - (u-\ft{i}{2})^{J-n}) + O(g^2)\, ,
\eeq
to leading order at weak coupling.

\subsection{Fixing the solution}

Now that we have fixed the $t$-function, we wish to fix the solution. We note that we already have a particular solution to the Baxter equation~(\ref{LBE}). Indeed, by construction, the function $\hat{r}(u)$, see Eq.~(\ref{rprim}), is solution if the relations~(\ref{qqrel}) are observed. But is this solution identical to the one, $r(u)$, we are looking for? This will be the case if the function $\hat{r}(u)$ is regular at small rapidity. This turns out not to be the case.

To see that $\hat{r}(u)$ is not regular at small rapidity we can consider the weak coupling limit, $g\sim 0$. We expect that in this limit the charges $q_{n+1}$ would be of order $O(g^0)$. Looking at the condition~(\ref{qqrel}) this implies that the charges
\beq
q_{-J-k+1} = O(g^{2k})\, ,
\eeq
are negligeable when $k\geq 1$. It means that the series representation~(\ref{rprim}) for the function $\hat{r}(u)$ truncates at weak coupling,
\beq\label{rprimWC}
\hat{r}(u) = i\sum_{n= 1}^{J}q_{-n+1}/u^n + O(g^2)\, .
\eeq
To leading order at weak coupling, the function $\hat{r}(u)$ is therefore singular at $u=0$. There is moreover no hope that the situation gets improved at higher orders in the weak coupling expansion. Perturbative corrections are actually more singular at small $u$.  

What we learn from this is that
\beq
r(u) \neq \hat{r}(u)\, .
\eeq
Nevertheless, both $r(u)$ and $\hat{r}(u)$ are solutions to the same linear equation~(\ref{LBE}). They should therefore differ by an homogenous solution, 
\beq\label{FSol}
r(u)-\hat{r}(u) = r_{hom}(u)\, ,
\eeq
where $r_{hom}(u)$ solves the homogeneous linearized Baxter equation,
\beq
x^{+ J}r_{hom}(u+\ft{i}{2}) - x^{- J}r_{hom}(u-\ft{i}{2}) = 0\, .
\eeq
The general solution to this equation is given by
\beq\label{rhom}
r_{hom}(u) = -i{f(u) \over x^J}\, ,
\eeq
where $f(u)$ is a $i$-periodic function, $f(u+i) = f(u)$.

We still have to impose the regularity condition~(\ref{regC}). Using the representation~(\ref{FSol}, \ref{rhom}) for $r(u)$, it yields
\beq\label{CReg}
q_{-k+1}-g^{2k}q_{k+1} = {1\over 2i\pi}\oint {dx \over x^{J+1}}\bigg[x^{k}-\left({g^2 \over x}\right)^k\bigg] f(u)\, .
\eeq
The left-hand side comes from the evaluation of the contour integral for $\hat{r}(u)$, as given in~(\ref{rprim}), with both terms obtained from the residue at $x=\infty$. We assumed therefore that the contour of integration could be deformed toward $x \sim \infty$. This is consitent with the fact that $\hat{r}(u)$ does not have singularities except at $u=0$, to any order in perturbation theory.

We can now obtain a more convenient representation for all the charges. We recall indeed that $q_{k+1}$ can be found from the expansion of $r(u)$ at small $u$, using Eq.~(\ref{Qru}). It leads to
\beq\label{SmC}
q_{k+1} = {1\over 2\pi}\oint {dx \over x^{k+1}}r(u) = {1\over 2i\pi}\oint {dx \over x^{J+k+1}}f(u)\, ,
\eeq
where we applied Eq.~(\ref{FSol}, \ref{rhom}), and evaluated the integral of $\hat{r}(u)$ by deforming the contour of integration to infinity as before, which then cancels exactly. Combining Eqs.~(\ref{SmC}, \ref{CReg}) we get also
\beq
q_{-k+1} = {1\over 2i\pi}\oint {dx \over x^{J-k+1}}f(u)\, .
\eeq
In other words, the representation~(\ref{SmC}) is equally valid for $k$ positive or negative. With its help, the condition~(\ref{qqrel}) is written as
\beq
{1\over 2i\pi}\oint {dx \over x}\bigg[x^{k} - {g^{2k} \over x^{k}}\bigg]f(u) = 0\, ,
\eeq
for $k\neq J$. In the following we shall also assume that the relation above holds for $k= J$. It implies that $f(u)$ is regular at $u=0$. Then we get $ q_{1} = q_{-2J+1}=0$ as the extra condition, meaning that we only consider zero momentum states.

\section{Small spin solution}

The main conclusion of the previous section is that the charges at small spin take the form
\beq
q_{n+1} = {1\over 2i\pi}\oint {dx \over x^{J+n+1}}f(u)\, ,
\eeq
with $n \in \mathbb{Z}$ and $f(u)$ a $i$-periodic function, regular around $u=0$. Now, to get more explicit expression, we shall make a specific choice for the function $f(u)$. 

\subsection{Elementary solution}

The simplest possible ansatz for the function $f(u)$, satisfying $f(u+i) = f(u)$, is certainly
\beq\label{ESol}
f(u) = C_{m} \, e^{2\pi mu}\, ,
\eeq
with $m \in \mathbb{Z}$ and $C_{m}$ a $u$-independent constant.  It defines what we call an elementary solution.

Let us see what are the charges associated to~(\ref{ESol}). Using the well-known formulae%
\footnote{Note that $I_{-k}(x) = I_{k}(x)$.}
\beq
{1\over 2i\pi}\oint {dx \over x^{k+1}} e^{ut} = {I_{k}(2gt) \over  g^k}\, , \qquad k \in \mathbb{Z}\, ,
\eeq
with $I_{k}(x)$ the $k$-th Bessel's function, we conclude that
\beq
q^{(m)}_{n+1} = {C_{m} \over g^{n+J}}I_{n+J}(4\pi m g)\, ,
\eeq
for $n \in \mathbb{Z}$. Here the upper index on the charge $q^{(m)}_{n+1}$ reminds us that it is obtained for the elementary solution~(\ref{ESol}).

We are half-way to the sought result~(\ref{ResInt}). We would like now to fix the arbitrary constant $C_{m}$. To do so, we require that the solution fulfills the condition~(\ref{Q0Q2}), which reads
\beq
q^{(m)}_{0} -g^2q^{(m)}_{2} = 1\, ,
\eeq
to leading order at small spin. It implies that
\beq
C_{m} = {2\pi m \over J}{g^{J} \over I_{J}(4\pi m g)}\, ,
\eeq
after using recurrence relation for the Bessel's functions. The final expression for the charges associated to the choice $f(u) = C_{m}\, e^{2\pi m u}$ is therefore
\beq
2g^{n+1}q^{(m)}_{n+1} = {m\sqrt{\lambda} \over J}{I_{n+J}(m \sqrt{\lambda}) \over I_{J}(m \sqrt{\lambda})}\, ,
\eeq
with $\sqrt{\lambda} \equiv 4\pi g$ and $n \in \mathbb{Z}$. This is the result announced previously in~(\ref{ResInt}). We notice that the arbitrary integer $m$ receives the meaning of a mode number, since for $n=0$ we get
\beq
q^{(m)}_{1} =  {2\pi m  \over J}\, .
\eeq
For a state with zero momentum, $m$ is forced to vanish. The only `on-shell' elementary solution is hence the zero-mode solution $m=0$. This solution is actually equivalent to $r(u)=0$. It is not identical to the vacuum solution however: It has $q_{-J+1} \neq 0$ with all other charges vanishing, and moreover $q_{-J+1}$ is arbitrary. This is an ambiguity of our construction. Since we do not know of any physical solution (in $\mathcal{N}=4$ SYM theory) involving zero mode, we discard it in the following.%
\footnote{An example of Bethe root $u$ filling the mode zero is $u=0$. It appears not easy to find solutions to the ABA equations, in the $\mathfrak{sl}(2)$ sector, with a root at the origin and the condition $\exp{(iQ_{1})}=1$ fulfilled at the same time.}

\subsection{General solution}

We did not find interesting solution carrying a single mode number $m$. Using the linearity of the Baxter equation at small spin, we can however easily combine elementary solutions with different mode numbers, by using the periodic function
\beq
f(u) = \sum_{m}\kappa_{m} C_{m} e^{2\pi m u}\, .
\eeq
Any such linear combination, with a given set of coefficients $\kappa_{m}$, is a new solution. The restriction to zero momentum states yields%
\footnote{We recall that this condition follows from $q_{-2J+1}=0$ (itself derived from the Baxter equation) when $f(u)$ is regular at $u=0$.} 
\beq
\sum_{m}{2\pi m \over J}N_{m} = 0 \,,
\eeq
with $N_{m}\equiv \kappa_{m}S$. From the (normalization) condition~(\ref{Q0Q2}) we also learn that
\beq
\sum_{m}N_{m} = S\, ,
\eeq
which means that the coefficient $\kappa_{m}$ is the filling fraction for the mode number $m$. It is then straighforward to derive the expression given in~\cite{Slope} for the minimal twist $J$ scaling dimension. The latter is well-known to be characterized by the condition that $\kappa_{1} = \kappa_{-1} =1/2$ with no other mode numbers filled. This immediately leads to~(\ref{slope}) and concludes our analysis.

\section{Concluding remarks}

In this paper, we have constructed a family of solutions to the long-range Baxter equation at small spin. It led in particular to a proof of an earlier proposal for the minimal scaling dimensions at small spin in the so-called $\mathfrak{sl}(2)$ sector. We stress that this derivation relies on the ABA equations only and therefore sheds no light on why wrapping effects appear subleading at small spin (in planar $\mathcal{N}=4$ SYM theory).

One embarrassing feature of our construction is that no restriction on the mode numbers and/or filling fractions was uncovered along the way. In other words, we found too many solutions. At twist two, for instance, there is only one physical trajectory, the one that interpolates between the (even-spin) scaling dimensions. In this case, we do not have the freedom of varying the filling fractions, neither can we fill mode numbers higher than $\pm 1$.%
\footnote{Higher twist solutions are slightly more flexible: The maximally allowed mode number is now $m=\pm (J-1)$ and the filling fractions are less constrained, but not entirely arbitrary.}
None of these constraints is visible within our approach. 

What is apparently missing is a sort of normalizability condition, which would trim the space of solutions we found. For illustration, the choice of the mode numbers obviously affects the large $u$ asymptotics of the solution, since
\beq
r(u) \sim e^{2\pi m u}\, ,
\eeq
with $m$ the largest mode number in the state. If a physical solution ought to be bounded, with respect to some norm, this could easily turn into a constraint on the maximally allowed mode number. It is less obvious, however, how one could put restrictions on the filling fractions themselves.

A more pessimistic conclusion is that the small spin expansion, which seems somehow meaningful when applied to the minimal scaling dimensions, does not make sense for higher trajectories. It is after all not guaranteed that an analytical extrapolation exists for these trajectories and even less certain that it would interpolate with the vacuum solution after continuation to zero spin. In this regard, it would be interesting to compare the strong-coupling predictions of our formula for higher trajectories to semiclassical string computations for the spiky strings~\cite{BRT}.

Finally, it would be nice to examine whether the method developed here could help in solving the recently proposed Boundary TBA equations for the cusp anomalous dimension~\cite{CMS, D} -- which is a wonderfully convoluted non-linear problem. The reason to believe such a relation could exist comes from the striking similarity between the formula~(\ref{slope}) and the one for the cusp anomalous dimension at small angle~\cite{CHMS, FGL}.

\textit{Note added:}

I was informed that an alternative derivation of the formulae presented in this paper was obtained by N.~Gromov~\cite{KG}.

\section*{Acknowledgements}

I would like to acknowledge to N.~Gromov for interesting communications and for sharing with me his results prior to publication. I am also most grateful to G.~Korchemsky for interesting discussions. 

\appendix

\section{Linearized Baxter equation}\label{App}

In this Appendix, we derive the linearized form of the Baxter equation and the conditions upon which the particular solution $\hat{r}(u)$, introduced in Section~3, solves this equation.

As recalled in Section~3, the long-range Baxter equation~\cite{Bel06, Bel09, BaBel} can be written as%
\footnote{We use the formulation given in~\cite{BaBel}, up to minor changes in the notations.}
\beq\label{Baxeq}
\Delta_{+}(u+\ft{i}{2})R(u + \ft{i}{2}) + \Delta_{-}(u-\ft{i}{2})R^{-1}(u - \ft{i}{2}) = t(u)\, ,
\eeq
for the ratio $R(u) = Q(u+i/2)/Q(u-i/2)$ of $Q$-functions. It involves the so-called dressing factors $\Delta_{\pm}(u)$, which are complex conjugate of one another, and the $t$-function $t(u)$, which is real. Both the dressing factors and $t$-function have complicated dependence on the spectral parameter $u$ and coupling constant. They are also dynamical, in the sense that they depend on the state under study, that is, on the solution $Q(u)$ of the Baxter equation. They simplify at weak coupling where $\Delta_{\pm}(u) \sim u^J$ and $t(u)$ reduces to a polynomial in $u$ of degree $J$. Starting from this input, they can be obtained iteratively, order by order, in the weak coupling expansion. In the following, we construct the expressions for all these quantities at small spin, but for arbitrary coupling.

\subsubsection*{Dressing factors}

We begin with the small spin expansion of the dressing factors $\Delta_{\pm}(u)$. They can be written in closed form as%
\footnote{The comparison with the notations $\gamma_{n}$ of~\cite{BaBel} yields
\beq
2\gamma_{2n-1} = (-1)^{n+1} g^{2n-1}\left(Q_{2n}-i\tilde{Q}_{2n}\right)\, , \qquad 2\gamma_{2n} = (-1)^{n}g^{2n}\left(\tilde{Q}_{2n+1}+iQ_{2n+1}\right)\, .
\eeq
In~\cite{BaBel}, the charges $Q_{n+1}, \tilde{Q}_{n+1}$ were denoted $q_{n+1}, \tilde{q}_{n+1}$, which we prefer to reserve here for their small spin expressions.}
\beq\label{AppDF}
\Delta_{\pm}(u) = x^{J}\exp{\bigg[\sum_{n\geq 1}\tilde{Q}_{n+1}\left({g^2 \over x}\right)^n \pm i \sum_{n\geq 1}Q_{n+1}\left({g^2 \over x}\right)^n \bigg]}\, .
\eeq
They are parameterized in terms of the charges $Q_{n+1}$, encountered in~(\ref{QMR}), and the secondary charges $\tilde{Q}_{n+1}$. In terms of the Bethe roots, the latter quantities stand for the moments
\beq\label{AuxC}
\tilde{Q}_{n+1} = {1\over n}\sum_{j=1}^{S}\bigg[{1\over x^{+ n}_{j}} + {1\over x^{-n}_{j}}\bigg] + \textrm{dressing phase corrections}\, ,
\eeq
which absorb all the contributions from the dressing phase. Their precise dependence on the dressing phase can be found in~\cite{BaBel} but is irrelevant here. It turns out indeed that the charges~(\ref{AuxC}) do not play any role in the construction of the small spin solution presented in Section~3. This `decoupling' between the charges $\tilde{Q}_{n+1}$ and $Q_{n+1}$ at small spin explains why there is no trace of the dressing phase in our result. 

At small spin, we can evaluate~(\ref{AppDF}) using the ansatz $\tilde{Q}_{n+1} = \tilde{q}_{n+1} S + O(S^2)$ and $Q_{n+1} = q_{n+1} S + O(S^2)$. It leads immediately to 
\beq\label{AppSDF}
\Delta_{\pm}(u) = x^{J}\bigg[1 + \delta_{\pm}(u)S + O(S^2)\bigg]\, ,
\eeq
with
\beq\label{deltapm}
\delta_{\pm }(u) = \sum_{n\geq 1}\tilde{q}_{n+1}\left({g^2 \over x}\right)^n \pm i\sum_{n\geq 1}q_{n+1}\left({g^2 \over x}\right)^n \, .
\eeq

\subsubsection*{$t$ function}

We now consider the $t$-function, which enters on the right-hand side of the Baxter equation~(\ref{Baxeq}). We can decompose it into two parts,
\beq
t(u) = p(u) + s(u)\, ,
\eeq
where $p(u)$ is a polynomial in $u$ of degree $J$, while $s(u)$ is non-polynomial and admits the expansion%
\footnote{Note that we have rescaled the expansion coefficients $s_{n}$ by a factor $1/2$ as compared to~\cite{BaBel}.}
\beq\label{su}
s(u) = \sum_{n\geq 1}s_{n}\left({ig \over x^+}\right)^n + \sum_{n \geq 1}\bar{s}_{n}\bigg({g \over ix^-}\bigg)^n\, .
\eeq
The coefficients $s_{n}, \bar{s}_{n}$ are complex conjugate of one another. A general expression for them was given in~\cite{BaBel}, whose form at small spin will be given shortly.

At small spin, we look for an expansion of the $t-$function as
\beq
t(u) = t_{\textrm{vac}}(u) + \delta t(u)S+ O(S^2)\, ,
\eeq
where $t_{\textrm{vac}}(u)$ is the vacuum (i.e., spin zero) $t$-function and $\delta t(u)$ the first deviation from it. Each function receives contribution from both the polynomial and non-polynomial component, i.e., $t_{\textrm{vac}}(u) \equiv p_{\textrm{vac}}(u) + s_{\textrm{vac}}(u)$ and $\delta t(u) \equiv \delta p(u) + \delta s(u)$.

We shall see below how to determine precisely $p_{\textrm{vac}}(u)$ and $\delta p(u)$. What we can say at the moment is that $\delta p(u)$ is a polynomial in $u$ of degree $J-1$. This follows from the fact that the leading large $u$ behavior of $p(u)$ is independent of the spin -- indeed $p(u) = 2u^J + O(u^{J-1})$ for any spin $S$ -- and thus present in $p_{\textrm{vac}}(u)$ only.

For the non-polynomial part of the $t$-function we can already obtain a more explicit expression. Using the formula~\cite{BaBel} for $s(u)$, we find to zero-th order at small spin that
\beq\label{svac}
s_{\textrm{vac}}(u) = -{g^{2J} \over x^{+J}} - {g^{2J} \over x^{-J}}\, .
\eeq
We stress that this result holds for physical states only, i.e., it assumes that $\exp{(iQ_{1})} = 1$.%
\footnote{In the notations of~\cite{BaBel} it means that $\exp{(i\vartheta)}=1$.}
At the next order, we find that $\delta s(u)$ admits representation as in~(\ref{su}) with $s_{n}$ replaced by
\beq\label{deltasn}
\delta s_{n} = {1\over 2i\pi}\oint {dx \over x}\left[\left({x\over ig}\right)^n - \bigg({g \over ix}\bigg)^n\right]x^{J}\bigg[-i\sum_{m\ge 1}q_{m+1}x^m + \sum_{m\ge 1}\tilde{q}_{m+1}\left({g^2 \over x}\right)^m\bigg]\, ,
\eeq
and with the integration taken counterclockwise around $x=0$.

The integral~(\ref{deltasn}) can be performed explicitely, providing
\beq
\delta s_{n} = i(-ig)^nq_{n-J+1} \theta(n-J) + (-i)^n g^{n+2J} \tilde{q}_{n+J+1}-(-i)^n g^{2J-n}\tilde{q}_{J-n+1}\theta(J-n)\, ,
\eeq
where $\theta(x) = 1$ if $x > 0$ and zero otherwise. Plugging these coefficients into the series representation~(\ref{su}) for $\delta s(u)$, we conclude that
\beq\label{deltasu}
\begin{aligned}
\delta &s(u) = \, \, i\sum_{n\geq 1}q_{n+1}\bigg[\left({g^2 \over x^+}\right)^{n+J}- \left({g^2 \over x^-}\right)^{n+J}\bigg] + \sum_{n\geq J+1}\tilde{q}_{n+1}\bigg[x^{+J}\left({g^2 \over x^{+}}\right)^n+ x^{-J}\left({g^2 \over x^{-}}\right)^{n}\bigg] \\
& \qquad \, \, \, \, - \sum_{n=1}^{J-1}g^{2n}\tilde{q}_{n+1}\bigg[\left({g^2 \over x^{+}}\right)^{J-n}+ \left({g^2 \over x^{-}}\right)^{J-n}\bigg]\, .
\end{aligned}
\eeq

\subsubsection*{Linearized Baxter equation}

We are now in position to derive the linearized Baxter equation. We recall that we look for a solution
\beq
R(u) = 1+r(u)S+ O(S^2)\, .
\eeq
Plugging this expansion into the Baxter equation~(\ref{Baxeq}), and using that
\beq
\Delta_{\pm}(u\pm \ft{i}{2})R^{\pm 1}(u\pm \ft{i}{2}) = x^{\pm J} + x^{\pm J}\bigg[\delta_{\pm}(u\pm \ft{i}{2}) \pm r(u\pm \ft{i}{2})\bigg]S + O(S^2)\, ,
\eeq
as follows from~(\ref{AppSDF}), we first observe that at zero-th order at small spin
\beq
x^{+J} + x^{-J} = t_{\textrm{vac}}(u) \, .
\eeq
We notice that at weak coupling it reduces to
\beq
t_{\textrm{vac}}(u) = (u+\ft{i}{2})^J + (u-\ft{i}{2})^J + O(g^2)\, ,
\eeq
since $x^{\pm} \sim u\pm i/2$. This is the right expression for the eigenvalue of the auxiliary transfer matrix of the XXX spin chain in its (ferromagnetic) vacuum.

Using our previous result~(\ref{svac}), we also conclude that the polynomial part of the $t$-function is given by
\beq\label{pvac}
p_{\textrm{vac}}(u) \equiv t_{\textrm{vac}}(u)-s_{\textrm{vac}}(u) = x^{+J} + {g^{2J} \over x^{+J}} + x^{-J} + {g^{2J} \over x^{-J}}\, .
\eeq
Though it is not transparent, the right-hand side of~(\ref{pvac}) is indeed a polynomial of degree $J$ in $u$. This follows from the fact that, more generally, the function
\beq\label{Un}
U^{(n)}(u) \equiv x^{n} + {g^{2n} \over x^{n}}\, ,
\eeq
that we shall encounter later on, defines a polynomial in $u$ of degree $n$. The function $U^{(n)}(u)$ is indeed regular at $u=0$, since
\beq
\oint {dx \over x}\bigg[x^{k} - \left({g^2 \over x}\right)^k\bigg]U^{(n)}(u) = 0\, ,
\eeq
for $k= 1, \ldots \,$. It thus defines an holomorphic function of $u\in \mathbb{C}$, which is simultaneously bounded by a polynomial at large $u$, $U^{(k)}(u) \sim u^J$. It is therefore polynomial itself, and so is
\beq
p_{\textrm{vac}}(u) = U^{(J)}(u+\ft{i}{2}) + U^{(J)}(u-\ft{i}{2})\, .
\eeq

If we now expand~(\ref{Baxeq}) up to the next order at small spin, we get the sought linearized Baxter equation. It reads
\beq\label{AppLin}
x^{+ J}r(u+\ft{i}{2}) - x^{-J}r(u-\ft{i}{2}) = t_{{\textrm{eff}}}(u)\, ,
\eeq
where, after combining everything together, the effective $t$-function $t_{{\textrm{eff}}}(u)$ is given by
\beq\label{Appteff}
t_{\textrm{eff}}(u) \equiv \delta p(u) +  \delta s(u) - x^{+J}\delta_{+}(u + \ft{i}{2}) - x^{-J}\delta_{-}(u-\ft{i}{2})\, .
\eeq

\subsubsection*{Particular solution}

We shall now require that the ansatz
\beq\label{rprime}
\hat{r}(u) = i\sum_{n\geq 1}q_{-n+1}\left({1\over x}\right)^n -i\sum_{n\geq 1}q_{n+1}\left({g^2 \over x}\right)^n\, ,
\eeq
solves the linearized Baxter equation~(\ref{Appteff}). Plugging this expression, and the one found for the effective $t$-function~(\ref{Appteff}, \ref{deltapm}, \ref{deltasu}), into the linearized Baxter equation~(\ref{AppLin}), we obtain, after a few algebra, the identity
\beq\label{LBEqF}
\begin{aligned}
\delta p(u)& -i\sum_{n = 1}^{J}q_{-n+1}\bigg[U^{+(J-n)} - U^{-(J-n)} \bigg]-\sum_{n=1}^{J}g^{2n}\tilde{q}_{n+1}\bigg[U^{+(J-n)} + U^{-(J-n)}\bigg] \\
&=\, \,  iq_{-2J+1}\bigg[{1\over x^{+J}}-{1\over x^{-J}}\bigg] + i\sum_{n\geq 1, n \neq J}(q_{-n-J+1}-g^{2n}q_{n-J+1})\bigg[{1\over x^{+n}}-{1 \over x^{-n}}\bigg] \, ,
\end{aligned}
\eeq
where we introduced the notation
\beq
U^{\pm (k)} \equiv U^{(k)}(u\pm \ft{i}{2}) = x^{\pm k} + {g^{2k} \over x^{\pm k}}\, ,
\eeq
with $U^{(k)}(u)$ the same polynomomial as in~(\ref{Un}). For $k=0$ we set $U^{\pm (0)} \equiv 1$.

The equation~(\ref{LBEqF}) is an identity between a polynomials in $u$, on the left-hand side, and a function suppressed at large $u$, on the right-hand side. Each term should therefore vanishes independently of the other. From the vanishing of the left-hand side of Eq.~(\ref{LBEqF}), we find that the polynomial $\delta p(u)$ is given by
\beq\label{deltap}
\delta p(u) = i\sum_{n=1}^{J}q_{-n+1}\bigg[U^{+ (J-n)} - U^{-(J-n)}\bigg] + \sum_{n=1}^{J}g^{2n}\tilde{q}_{n+1}\bigg[U^{+ (J-n)} + U^{-(J-n)}\bigg] \, .
\eeq
At weak coupling, it simplifies to
\beq
\delta p(u) = i\sum_{n=1}^{J}q_{-n+1}\bigg[(u+\ft{i}{2})^{J-n} - \left(u-\ft{i}{2}\right)^{J-n}\bigg] + O(g^2)\, ,
\eeq
since all the charges $\tilde{q}_{n+1}$ are expected to be of order $O(g^0)$. Finally, from the vanishing of the right-hand side of Eq.~(\ref{LBEqF}), we get
\beq
q_{-n-J+1} = g^{2n}q_{n-J+1}\, ,
\eeq
for $n\neq J$, and $q_{-2J+1} = 0$ otherwise.

\end{document}